\documentclass[pre,twocolumn,showpacs,superscriptaddress,aps]{revtex4}
\usepackage[dvipdfmx]{graphicx}
\usepackage{amsfonts,xfrac}
\usepackage{amsmath}
\usepackage{amssymb}
\usepackage{graphicx}
\usepackage{dcolumn}
\usepackage{dsfont}
\usepackage{times}
\usepackage{epsfig,color}
\usepackage[]{units}
\usepackage{soul}
\usepackage{bm}
\usepackage[hang,nooneline]{subfigure}                         
\newcounter{fig}

\def\epsilon{\varepsilon}

\begin{document}
\title{Shock and Rarefaction Waves in Generalized Hertzian Contact Models}

\author{H. Yasuda}
\affiliation{Aeronautics \& Astronautics, University of Washington, Seattle, WA 98195-2400, USA}
\author{C. Chong}
\affiliation{Department of Mathematics, Bowdoin College, Brunswick, ME 04011, USA}
\author{J. Yang}
\affiliation{Aeronautics \& Astronautics, University of Washington, Seattle, WA 98195-2400, USA}
\author{P. G. Kevrekidis}
\affiliation{Department of Mathematics and Statistics, University of Massachusetts, Amherst, MA 01003-4515, USA}

\date{\today}

\begin{abstract}
  In the present work motivated by generalized forms of the Hertzian
  dynamics associated with granular crystals, we consider the possibility
  of such models to give rise to both shock and rarefaction waves.
  Depending on the value $p$ of the nonlinearity exponent, we find
  that both of these
  possibilities are realizable. We use a quasi-continuum approximation
  of a generalized inviscid Burgers model in order to predict the solution
  profile up to times near the shock formation, as well as to estimate
  when it will occur. Beyond that time threshold, oscillations associated
  with the discrete nature of the underlying model emerge that cannot
  be captured by the quasi-continuum approximation. Our analytical
  characterization of the above features is complemented by systematic
  numerical computations.
\end{abstract}

\maketitle

\section{Introduction}
Over the last two decades, the examination of granular crystals
has received considerable attention, as is now summarized in a
wide range of reviews~\cite{Nester2001,sen08,theocharis_review,ptpaper,vakakis_review}.
Granular crystals consist of closely packed arrays of particles typically modeled as interacting with each other elastically
via so-called Hertzian contacts.
 The resulting force depends on the geometry of the particles,
the contact angle, and 
elastic properties of the particles~\cite{Johnson}. Part of the
reason for the wide appeal of these systems is associated with their
remarkable tunability, which permits to access weakly (via application
of a prestress/precompression and for strains considerably lower
than the precompression) or strongly (in the absence of such precompression)
nonlinear regimes. At the same time, it is possible to easily access
and arrange the media in homogeneous or heterogeneous configurations.
Notable examples of the later include periodically arranged chains or those involving disorder.
It is, thus, not surprising that granular crystals have been explored as a prototypical playground
for a variety of applications including, among others,
shock and energy absorbing 
layers~\cite{dar06,hong05,fernando,doney06}, actuating devices \cite{dev08}, acoustic lenses \cite{Spadoni}, acoustic diodes \cite{Nature11} and switches \cite{Li_switch}, and sound scramblers \cite{dar05,dar05b}.

Much attention has been paid to special nonlinear solutions of the granular chain.
Perhaps the most prototypical 
example explored in the context of granular crystals (with or without precompression)
is the traveling solitary wave~\cite{Nester2001,sen08}. More recently, 
both theoretical/computational and experimental interest has been
expanded to another broad class of solutions, so-called bright and
dark discrete breathers, which are exponentially
localized in space and periodic in time~\cite{theocharis_review,ptpaper}. 

Our aim in the present work is to revisit a far less explored family
of waveforms, namely shock waves. Shock waves are characterized by an abrupt, 
nearly discontinuous change in the wave \cite{scholar}. In the context of partial 
differential equations (PDEs), such as the inviscid Burgers equations,
this definition can be made more precise to be a solution that develops
infinite derivatives in finite time \cite{Smoller}.
Shock-like structures have been studied experimentally in homogeneous
granular chains in the absence of precompression~\cite{herbold,Molinari2009}.
In those works, a shock wave was generated by applying a velocity to a
single particle~\cite{herbold} or by imparting velocity continuously to the end of a chain~\cite{Molinari2009}.
More generally, in the context of the celebrated Fermi-Pasta-Ulam (FPU)
lattices (of which the granular chain is a specific example of), dispersive shock waves were examined numerically for the
case of general convex FPU potentials for arbitrary Riemann (i.e.,
jump) initial data~\cite{Herrmann10}.

In this study, we generalize the approach
taken in the work of~\cite{pikovsky}. There, it was recognized
that the quasi-continuum form of the Hertzian model directly
relates to a second-order PDE; 
for a rigorous justification, see~\cite{dcdsa}. Considering then the first-order nonlinear
transport PDEs that represent the right and left moving waves,
one retrieves effective models of the generalized family of
the inviscid Burgers type~\cite{mcdonald}. This enables the use of characteristics
in order to predict the evolution of initial data, as well as
the potential formation of shock waves.


The most commonly studied granular crystals are those consisting of strain-hardening materials. 
This implies that the nonlinear exponent satisfies $p>1$ in $F \propto \delta^p$, where $F$ and $\delta$ are compressive force and displacement.
For example, in the Hertzian case of spherical particles, the nonlinear exponent is $p=3/2$ \cite{Hertz}.
More recently, generalizations of the Hertzian contact law have been explored
in the context of mechanical metamaterials, including those with strain-softening behavior (where $0<p<1$) \cite{RarefactionNester}.
Examples of this include tensegrity structures~\cite{carpe}
and origami metamaterial lattices~\cite{hiromi1,origami}.
Motivated by those recent works,
we will examine the formation of shock waves for general values of $p$.
We find a fundamentally different dynamical behavior for the two cases of $p>1$ and $p<1$.
For $p>1$ where,
for monotonically decreasing initial data, a shock forms due to the larger
amplitudes traveling faster from the smaller ones. In the case of $p<1$,
the same initial data lead to a rarefaction waveform, where parts of the wave with small amplitude travel faster than larger amplitude parts of the wave \cite{strauss}.
On time scales where the quasi-continuum approximation remains valid, we are able
to analytically follow the solutions for both strain-hardening ($p>1$) and strain-softening materials ($p<1$). In the vicinity of the
shock time (which we predict in reasonable agreement with the
numerics from the quasi-continuum approximation), the discrete nature
of the underlying lattice emerges and significantly affects the dynamics.
Our systematic simulations illustrate -- by varying quantities like the
precompression or the nonlinearity power -- how these predictions become
progreessively less acurate as the model deviates from its linear analogue.

Our presentation of the above results will be structured as follows.
In section II, we will provide the theoretical background and analytical
findings associated with the generalized model. In section III,
we will compare theoretical predictions to direct numerical simulations of the particle
model. Finally, in section IV, we summarize our findings and
present some possible directions for future work.

\section{Theoretical Analysis}
Let $u_n$ be the relative displacement from equilibrium of the $n$-th particle in the lattice.
As discussed in the introduction, we are motivated by the generalization of the
Hertzian contact model to arbitrary powers not only due to
applications~\cite{hiromi1,origami,carpe}, but also due to significant
theoretical developments (e.g., regarding traveling waves)~\cite{pikovsky,pelingj,Dumas}. In that light, we 
consider the following equations of motion:
\begin{equation} \label{eq:EquationMotion}
	{{\ddot{u}}_{n}}=\left[ {{\delta }_{0}}+{{u}_{n-1}}-{{u}_{n}} \right]_{+}^{p}-\left[ {{\delta }_{0}}+{{u}_{n}}-{{u}_{n+1}} \right]_{+}^{p},
\end{equation}
where $\delta_0$ is the precompression (static load) applied to the system and the overdot represents
differentiation with respect to normalized time.
Defining the strain as ${{y}_{n}}={{u}_{n-1}}-{{u}_{n}}$, we rewrite Eq.~\eqref{eq:EquationMotion} as
\begin{equation}  \label{eq:EquationMotion2}
  {{\ddot{y}}_{n}}=
  {{\left( {{\delta }_{0}}+{{y}_{n-1}} \right)}^{p}}+{{\left( {{\delta }_{0}}+{{y}_{n+1}} \right)}^{p}}  
  -2{{\left( {{\delta }_{0}}+{{y}_{n}} \right)}^{p}}.
\end{equation}
Following~\cite{dcdsa} towards the derivation
of a long-wavelength approximation characterized
by smallness parameter ($\varepsilon$) and associated spatial
and temporal scales, respectively,
$X=\varepsilon n$ and $T=\varepsilon t$,
we express the strain as ${{y}_{n}}(t)=Y\left( X,T \right)$
and obtain:
\begin{eqnarray*}
	{{\varepsilon }^{2}}\partial _{T}^{2}Y=&&{{\left\{ {{\delta }_{0}}+Y\left( X+\varepsilon  \right) \right\}}^{p}}+{{\left\{ {{\delta }_{0}}+Y\left( X-\varepsilon  \right) \right\}}^{p}}  \\ &&-2{{\left\{ {{\delta }_{0}}+Y(X) \right\}}^{p}}.
\end{eqnarray*}
Hence, the continuum model follows:
\begin{equation}
  \partial _{T}^{2}Y=\partial _{X}^{2}\left\{ {{\left( {{\delta }_{0}}+Y \right)}^{p}} \right\}.
  \label{mcdon}
\end{equation}
Now, in the spirit of~\cite{mcdonald}, consider the first order
(nonlinear transport) PDEs that would be compatible with Eq.~(\ref{mcdon}).
\begin{equation} \label{eq:1st_orderPDE}
	{{\partial }_{T}}Y\pm \alpha {{\partial }_{X}}\left\{ {{\left( {{\delta }_{0}}+Y \right)}^{c}} \right\}=0,
\end{equation}
where $\pm$ indicates the two propagation directions. The parameters $\alpha$ and $c$ will be chosen
such that solutions of Eq.~\eqref{eq:1st_orderPDE} are solutions of Eq.~\eqref{mcdon}.
From this we infer:
\begin{equation}  \label{eq:2nd_orderPDE}
	\partial _{T}^{2}Y=  {{\partial }_{T}}\left[ \mp \alpha {{\partial }_{X}}\left\{ {{\left( {{\delta }_{0}}+Y \right)}^{c}} \right\} \right]=  \mp \frac{\alpha^2 {{c}^{2}}}{2c-1}\partial _{X}^{2}\left\{ {{\left( {{\delta }_{0}}+Y \right)}^{2c-1}} \right\}.
\end{equation}
Comparing Eq.~\eqref{mcdon} and the right hand side of the last equality of Eq.~\eqref{eq:2nd_orderPDE}, we obtain $c=(p+1)/2$
and $\alpha^2 = \frac{2c-1}{c^2}$. Using these definitions of $c$ and $\alpha$, we re-write
Eq.~\eqref{eq:1st_orderPDE} as
\begin{equation} \label{eq:BurgersEq}
	{{\partial }_{T}}Y \pm \sqrt{p} {{\left( {{\delta }_{0}}+Y \right)}^{\frac{p-1}{2}}}{{\partial }_{X}}Y=0.
\end{equation}
Using then the standard technique of characteristics (see, e.g.,~\cite{strauss}
for an elementary discussion) in order
to solve Eq.~(\ref{eq:BurgersEq}), we obtain the equation along
characteristic lines (along which the solution is constant)
of the form:
\begin{equation} \label{eq:CharacteristicEq}
	\frac{dX}{dT}=\phi \left( Y\left( X,T \right) \right)
\end{equation}
where
\begin{equation} \label{eq:Function}
		\phi \left( Y\left( X,T \right) \right)=\sqrt{p} \,  {\left( {{\delta }_{0}}+Y\left( X,T \right) \right)}^{\frac{p-1}{2}}.
\end{equation}
 Since solutions of Eq.~\eqref{eq:BurgersEq} are constant along characteristic lines, Eq.~\eqref{eq:CharacteristicEq} becomes
\begin{equation} \label{eq:Characteristics}
	\frac{dX}{dT}=\frac{X-{{X}_{0}}}{T}=\phi \left( Y\left( X,T \right) \right)=\phi \left( Y\left( {{X}_{0}},0 \right) \right).
\end{equation}
In this study, we choose
\begin{equation} \label{eq:InitialCondition}
	Y_0\left( {{X}_{0}} \right)=a\text{sech}\left( b{{X}_{0}} \right)
\end{equation}
as the initial function, motivated by the interest in localized initial
data.

As is well known in this broad class of generalized
inviscid Burgers models, despite the smooth initial
distribution of the strain ($y_n$), a shock wave (a wave breaking effect)
can be observed to form in a finite time; i.e., the solution develops
infinite derivatives in a finite time in the continuum limit of the
problem \cite{Smoller}.
This shock wave is created (due to the resulting multi-valuedness)
when the characteristic lines intersect.
 To predict when the shock wave is formed, i.e., the shock wave time ($T_s$), we consider the following two (arbitrary) characteristic lines.
\begin{eqnarray}
   X(T)&=&\phi \left( {{X}_{1}} \right)T+{{X}_{1}} \\
   X(T)&=&\phi \left( {{X}_{2}} \right)T+{{X}_{2}}
\end{eqnarray}
where ${{X}_{1}},{{X}_{2}}\in \mathbb{R}$ and $X_2=X_1+h$.
When these two lines intersect, we obtain
\begin{equation}
	T=-\frac{{{X}_{2}}-{{X}_{1}}}{\phi \left( {{X}_{2}} \right)-\phi \left( {{X}_{1}} \right)}=-\frac{h}{\phi \left( {{X}_{1}}+h \right)-\phi \left( {{X}_{1}} \right)}.
\end{equation}
The shock wave is formed at the time
at which the characteristic lines intersect for the first time.
Therefore, the shock time is calculated as follows:
\begin{eqnarray*}
   {{T}_{s}}&=&\min \left[ -\frac{h}{\phi \left( {{X}_{1}}+h \right)-\phi \left( {{X}_{1}} \right)} \right] \\ 
            &=&\frac{1}{\min \left[ -\frac{\phi \left( {{X}_{1}}+h \right)-\phi \left( {{X}_{1}} \right)}{h} \right]}.
\end{eqnarray*}
Then, considering $h\to 0$, we obtain
\begin{equation}
	{{T}_{s}} = -\frac{1}{\min \left[ \frac{d\phi (X)}{dX} \right]}.
\label{prediction}
\end{equation}
In this study, our semi-analytical prediction based on the
above considerations consists of evaluating $\frac{d\phi (X)}{dX}$
and (numerically) identifying the relevant minimuum, which leads via
Eq.~(\ref{prediction}) to a concrete estimate of $T_s$ to be compared
with direct numerical simulations.

\section{Numerical Computations and Comparison}

Armed with the above theoretical considerations, we now turn to a comparison
of the discrete model and the continuum model in terms of the formation of
shock waves (as a way of quantifying the accuracy of our prediction).
In order to initialize the direct simulation of the discrete model,
the initial velocity of each particle is needed.
Based on the initial strain $Y_0(X)$, the initial velocity can be computed as 
\begin{eqnarray*}
	\left[ \frac{d{{y}_{n}}}{dt} \right]_{t=0}  &=& \frac{d}{dt} \left[   Y(X,T)    \right]_{t=0} \\
	&=& \pm \epsilon \left[ \sqrt{p} {{\left( {{\delta }_{0}}+Y_0 \right)}^{\frac{p-1}{2}}}{{\partial }_{X}}Y_0 \right].
\end{eqnarray*}
The initial strain is given by Eq.~\eqref{eq:InitialCondition} with the numerical constants $a=0.01$ and $b=0.3$.
These simulations will be compared to the continuum model approximations.
The spatial profile of the continuum approximation at a given time $t$ can be computed by solving the following implicit equation,
\begin{equation} \label{cont_profile}
  Y(\epsilon n, \epsilon t) = Y_0( \epsilon n - \phi \epsilon t      ) , 
\end{equation}
where the wave velocity $\phi =  \pm \sqrt{p(\delta_0 + Y(\epsilon n, \epsilon t) )^{p-1}  }$ follows directly from Eq. \eqref{eq:BurgersEq}.

\begin{figure}[htbp]
\centerline{ \includegraphics[width=0.5\textwidth]{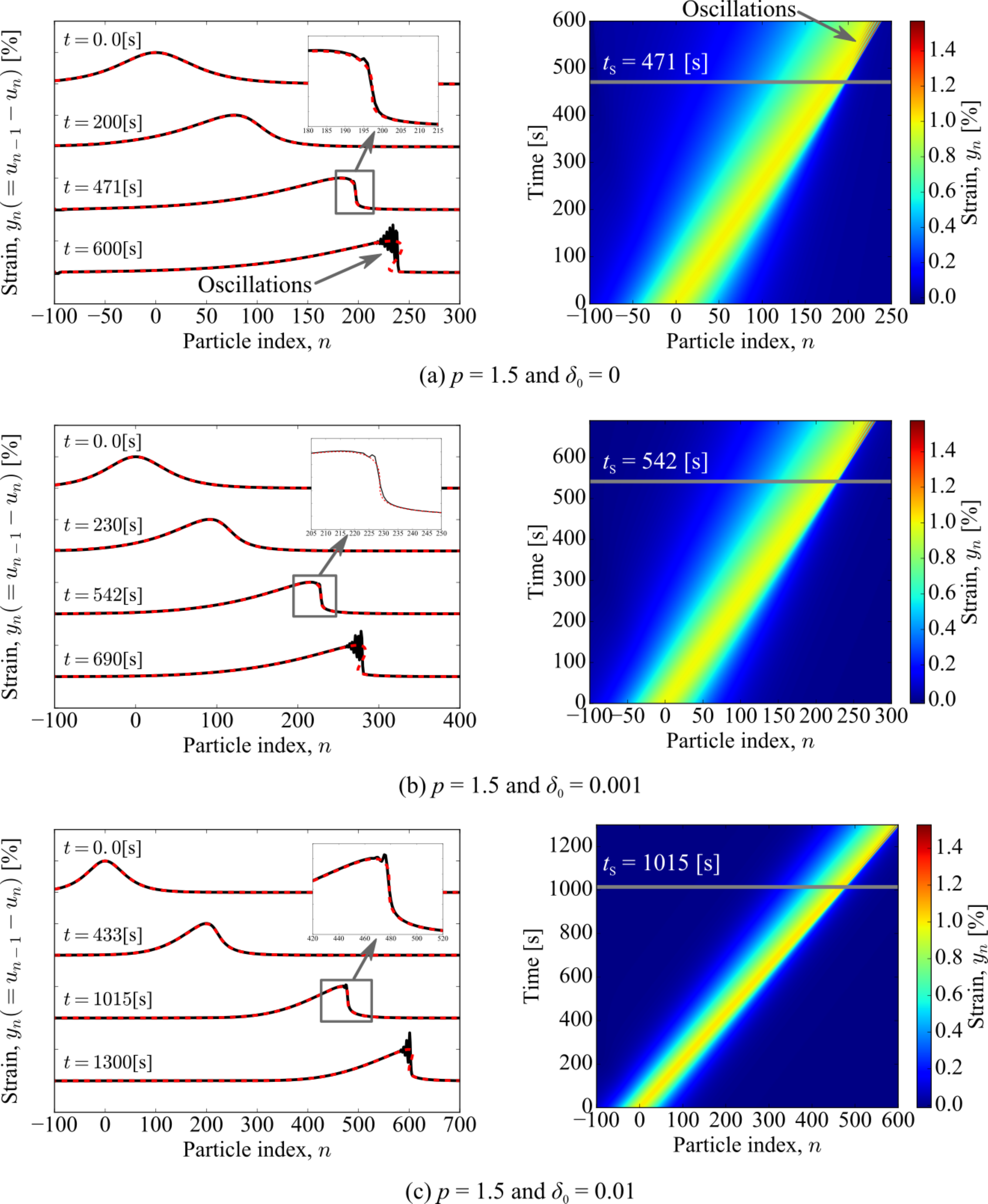}}
\caption{Strain wave propagation for $p=1.5$ and \textbf{(a)} $\delta_0=0$, \textbf{(b)} 0.001, and \textbf{(c)} 0.01. Left figures show temporal plots of strain waves. The black solid line is obtained from the direct simulation of the discrete model, and the red dashed line is the prediction based on Eq.~\eqref{cont_profile} (i.e., the continuum model). Strain curves are offset to ease visualization (ticks in the vertical axis indicate 1.0). The inset shows the magnified view of the leading part at the predicted shock time. The right figures show space-time contour plots of strain wave propagation. The gray solid line indicates the predicted analytical shock wave time.}
\label{fig:Shock_p15}
\end{figure}

\begin{figure}[htbp]
\centerline{ \includegraphics[width=0.5\textwidth]{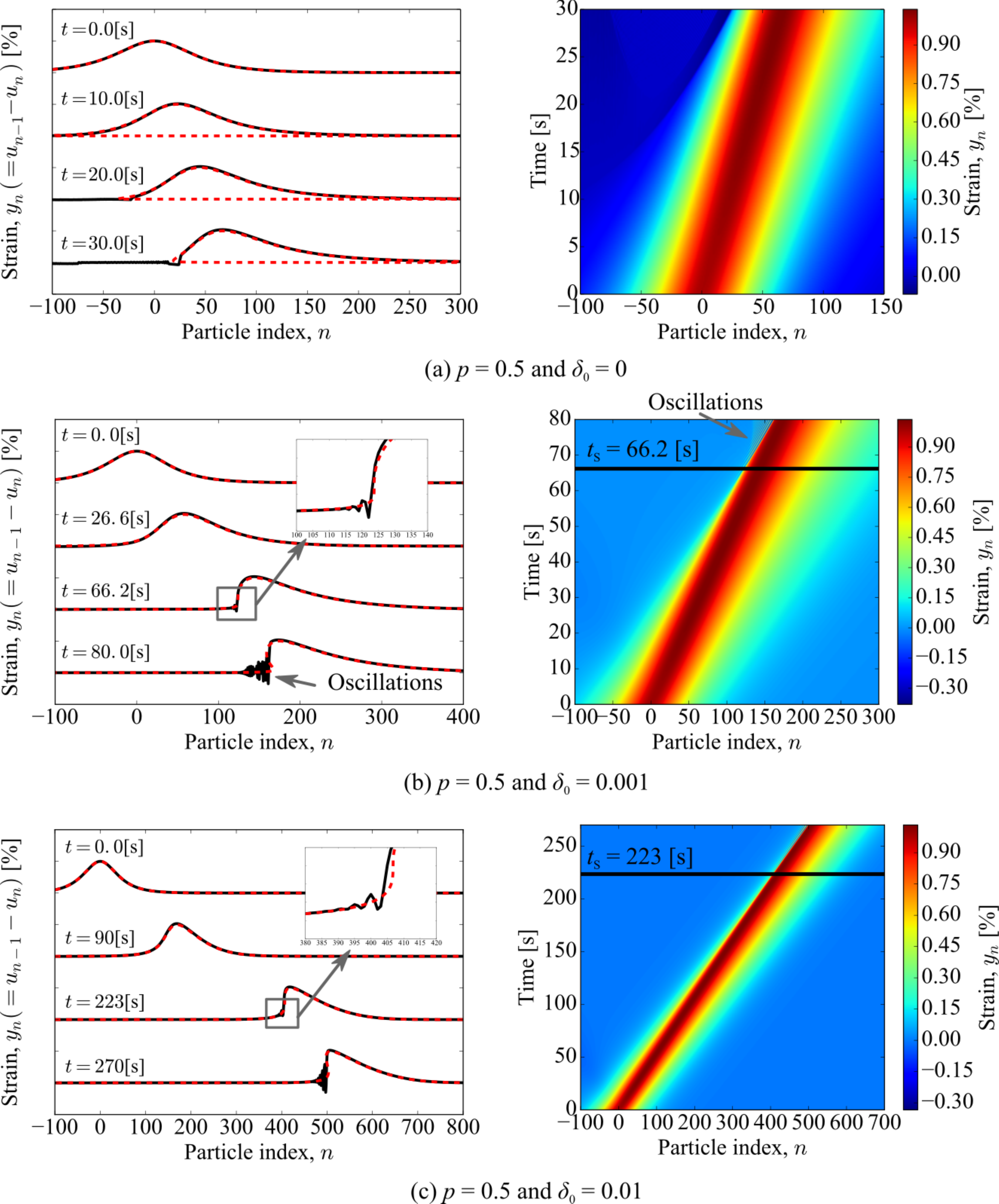}}
\caption{Strain wave propagation for $p=0.5$ and \textbf{(a)} $\delta_0=0$, \textbf{(b)} 0.001, and \textbf{(c)} 0.01. Left figures show temporal plots of strain waves. The black solid line is obtained from the direct simulation of the discrete model, and the red dashed line is the prediction based on Eq.~\eqref{cont_profile} (i.e., the continuum model). Strain curves are offset to ease visualization (ticks in the vertical axis indicate 1.0). The inset shows the magnified view of the leading part at the predicted shock time. The right figures show space-time contour plots of strain wave propagation. The gray solid line indicates the predicted analytical shock wave time.}
\label{fig:Shock_p05}
\end{figure}

 Figure~\ref{fig:Shock_p15} shows the strain wave propagation for
 the standard case of spherical Hertzian contacts
 $p=1.5$~\cite{Nester2001,sen08}, while Fig.~\ref{fig:Shock_p05} motivated
 by the works of~\cite{hiromi1,origami,carpe} shows the case of $p=0.5$.
 In these two figures, the left panels show the wave shape at four different
 time instances, and the right panels show the space-time contour plots of strain wave propagation.
 In each case the top row represents the case of no-precompression
 (fully nonlinear case), while progressively the middle and bottom
 row introduce higher precompression rendering the problem
 progressively more nonlinear. 
 For $p=1.5$, the wave breaks from its front part (see Fig.~\ref{fig:Shock_p15}), whereas the $p=0.5$ case shows the wave breaking from the tail part as shown in Fig.~\ref{fig:Shock_p05}. This can be understood since
 the speed of propagation $\phi(Y(X,T))$ in the system depends on the amplitude of the wave, see Eq.~\eqref{eq:Function}.
For $p>1$, points along the initial condition with larger amplitudes travel faster than points with smaller amplitudes.
 Thus, the large amplitude part of the wave overtakes the smaller amplitude part,
 leading to a shock formation in the front (monotonically
 decreasing part of the data), and a rarefaction in the back
 (monotonically increasing part of the data). On the contrary,
 for $p<1$, the situation gets reversed as is natural to infer
 from the speed expression $\phi(Y(X,T))$:  the rarefaction forms in the front,
 while the wave breaking emerges in the back. 

 In Figs.~\ref{fig:Shock_p15} and~\ref{fig:Shock_p05}, in addition
 to showing the dynamics of the original underlying lattice of
 Eq.~(\ref{eq:EquationMotion}), the prediction based on the theoretical
 analysis of the nonlinear transport PDE is shown, see Eq.~\eqref{cont_profile}.
 It can be inferred that the two closely match {\it until} the time where
 a discontinuity is about to form at which time the discrete
 model starts to feature oscillatory dynamics, a dispersive feature
 absent in the continuum, long wavelength model. 
 A specific diagnostic in that connection that we use in order to compare
 the discrete and continuum cases is the shock formation time $T_s$. 
 To obtain this time from the simulations, we examine the slope of the wave
 shape by calculating the sign of the differences of the strains between adjacent particles, and we define the shock wave time ($T_S$) when the sign changes multiple times, i.e., when the discrete character of the model prevents
 (the continuum)
 shock wave formation.
 The resulting comparison of $T_s$ is a principal quantitative finding
 of the present work. 

\begin{figure}[htbp]
\centerline{ \includegraphics[width=0.5\textwidth]{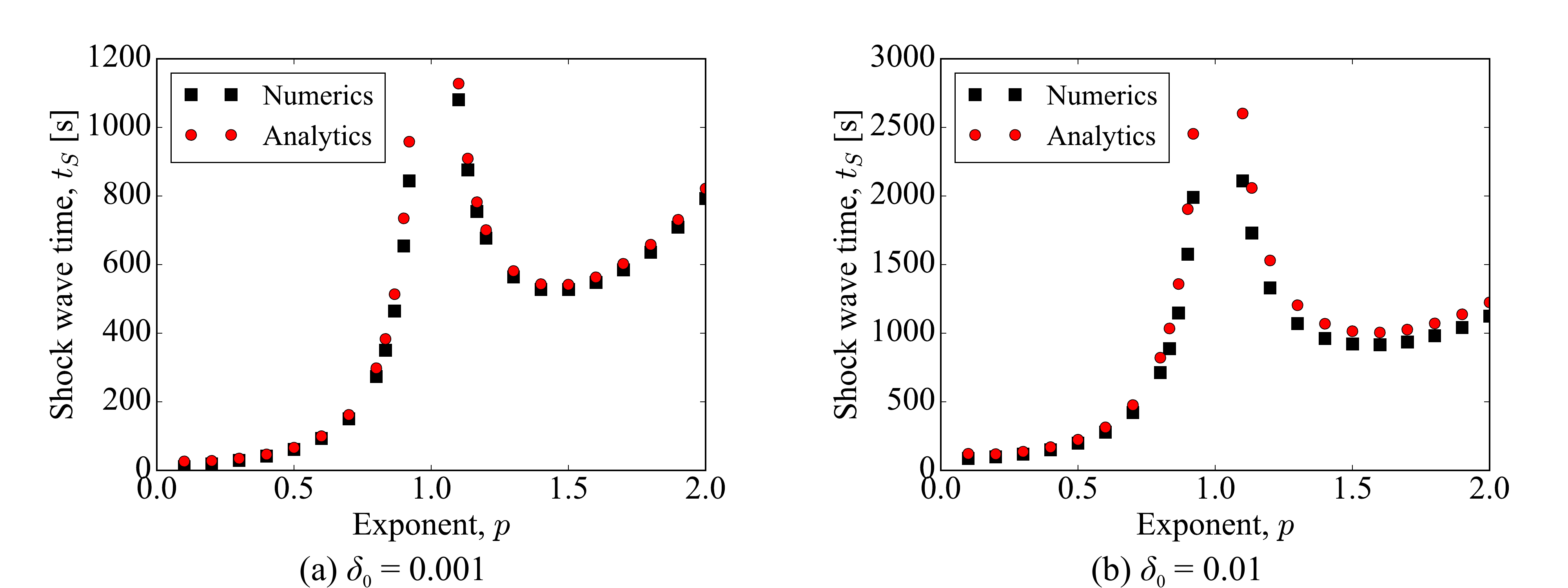}}
\caption{Effect of the exponent value ($p$) on the shock wave time for \textbf{(a)} $\delta_0=0.001$ and \textbf{(b)} $\delta_0=0.01$. The black square indicates the shock wave time from numerical simulations ($T_S$), and the red circle is the analytical prediction ($T_s$).}
\label{fig:Exponent_effect}
\end{figure}

\begin{figure}[htbp]
\centerline{ \includegraphics[width=0.5\textwidth]{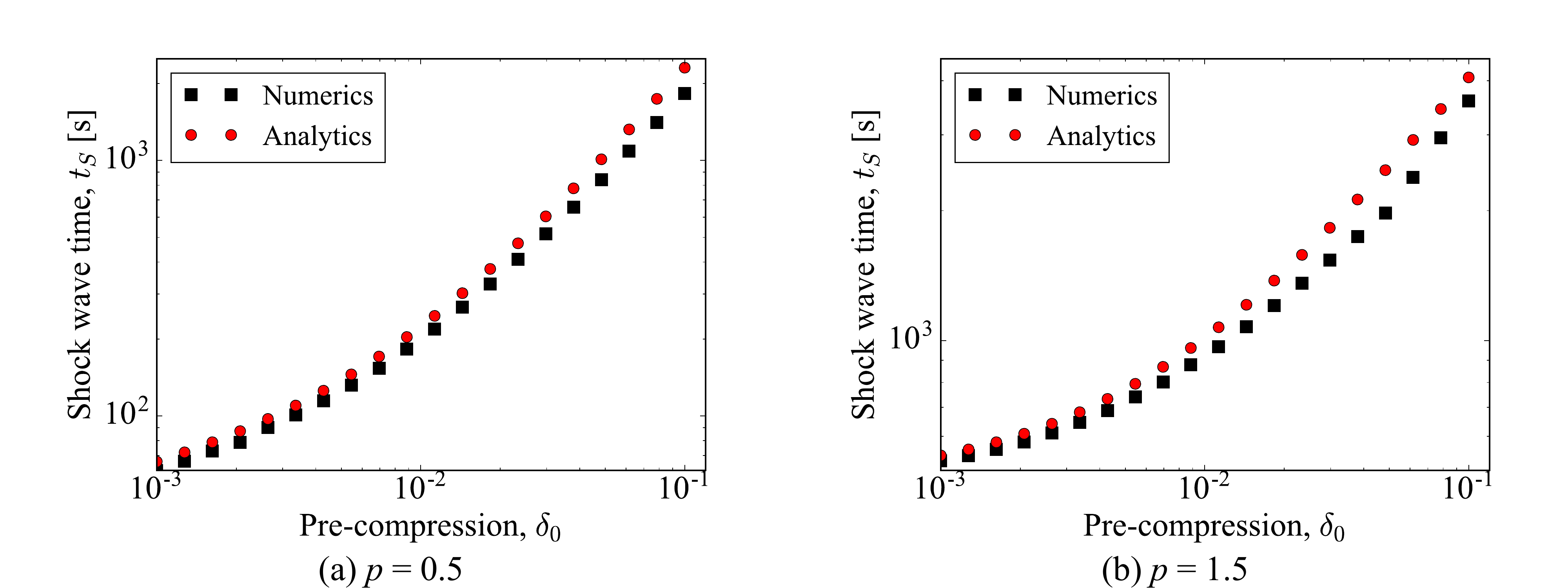}}
\caption{ 
Effect of the pre-compression ($\delta_0$) on the shock wave time for \textbf{(a)} $p=0.5$ and \textbf{(b)} $p=1.5$. The black square indicates the shock wave time from numerical simulations ($T_S$), and the red circle is the analytical prediction ($T_s$). Data is plotted on log-log scale.}
\label{fig:Precompression_effect}
\end{figure}

 Figure~\ref{fig:Exponent_effect} shows the comparison between numerical and analytical shock wave time for different exponent values ($p$).
 The difference increases if $p$ approaches unity
 because $p=1$ indicates that the system becomes a linear advection equation which does not support shock waves. It is for that reason that the shock formation
 time diverges as the limit $p \rightarrow 1$ is approached.
 Nevertheless, the quantitative trend between the two cases
 ($p>1$ and $p<1$) is well captured in both panels of the figure,
 i.e., for different values of the nonlinearity. In addition, we analyze the effect of pre-compression ($\delta_0$) on the shock wave time as shown in Fig.~\ref{fig:Precompression_effect}.
 As we increase the pre-compression (i.e., the system enters
 the weakly nonlinear regime and progressively
 approaches linear regime), the difference between numerical and analytical
 results increases.
 One can argue that this disparity may be partially
 related to the way we quantify
 $T_s$. In particular, as the system approaches $p=1$ or $\delta_0$ large,
 linear modes become progressively more accessible to it (a situation to
 be contrasted with the sonic vacuum where no such modes exist).
 As a result, while in the sonic vacuum -- in the effective absence of
 linear modes -- the time of the shock nearly coincides with the
 emergence of non-monotonicity near the wave front, in the progressively
 nearer linear case, the non-monotonicity emerges earlier. Hence the
 numerical evaluation of $T_s$ (based on the emergence of
 discreteness induced oscillations) suggests that $T_s$ is calculated to be
 well before the wave breaking event predicted by the analytics.
 Despite this distinction, it is fair to say that the quasi-continuum
 theory provides a very good tool for predicting the pre-shock
 wavefront evolution and for estimating the shock time, except in the
 vicinity of linear dynamical evolution.
 
\section{Conclusions \& Future Challenges}


 In the present work, we have investigated shock and rarefaction
 wave formation in generalized granular crystal systems with Hertzian
 contacts. Motivated by recent developments in origami
 metamaterials~\cite{hiromi1,origami} and tensegrity structures~\cite{carpe},
 we explored both the scenario of $p>1$ (including $p=3/2$ of spherical
 contacts), and that of $p<1$ of strain-softening media. We identified
 the emergence of both shock waves and rarefaction waves and found
 a complementarity between the two cases. In the strain-hardening
 case, shock waves arise out of monotonically decreasing initial conditions
 while rarefactions out of monotonically
 increasing ones. The reverse
 occurs in the case of strain-softening $p<1$ media. 
 The generalization of the quasi-continuum formulation
 of~\cite{dcdsa} in the spirit of the nonlinear transport equations
 proposed by~\cite{mcdonald}
 provided us with an analytical handle in order to characterize
 the evolution of pulse-like data in the strain variables. 
 A key quantitative prediction concerned the time of formation
 of the discontinuity as characterized at the quasi-continuum level.
 We discussed how the discreteness of the system,
 once visiting scales comparable
 to the lattice spacing, \textit{intervenes} by generating oscillations and
 departing in this way from the quasi-continuum description.

 This work suggests a number of exciting possibilities for the future.
 One interesting avenue is that of obtaining information about dispersive
 shock waves either at the level of the KdV model, or at that of the
 Toda lattice and then. Then, in the spirit of~\cite{Shen}, we can use these
 to characterize the formation of dispersive shock waves that
 emerge in the presence of precompression. That being said,
 it is unclear what becomes of such dispersive shock waves
 when precompression is weak or absent, as especially in the latter
 case there is no definitive characterization of dispersive shock waves.
 However, the direction of the first order equations could be an
 especially profitable one in that context. The work of~\cite{wilma}
 (see also numerous important references therein) suggests that for
 the inviscid Burgers problem shock waves have been extensively studied
 both analytically and numerically. It is conceivable that a
 (judiciously selected) discretization of the first order PDEs considered
 herein could offer considerable insight in the formation of shock
 waves, both in the weak and even in the case of vanishing precompression
 limit. These are important directions that are under current consideration
 and will be reported in future publications.

\begin{acknowledgments}
H.Y. and J.Y. acknowledge the support of the NSF under Grant No. 1553202.
The work of C.C. was partially funded by the NSF under Grant No. DMS-1615037. 
P.G.K gratefully acknowledges the support of the ERC under FP7; Marie
Curie Actions, People, International Research Staff Exchange
Scheme (IRSES-605096), and the Alexander von Humboldt Foundation.
J.Y. and P.G.K. are grateful for the support of the ARO under grant W911NF-15-1-0604.
\end{acknowledgments}





\bibliography{granular2}

\end{document}